\documentclass{aa}
\usepackage{graphics,amsmath}

\begin{document}
\title{Kinematic model inversions of hot star recurrent DAC data -- tests against
dynamical CIR models}
\titlerunning{Tests of kinematic model inversions of DAC data against dynamical CIR models}

\author{J.  Krti\v{c}ka\inst{1,2,3} \and 
R. K. Barrett\inst{1} \and   J. C. Brown\inst{1} \and S. P. Owocki\inst{1,4}}

\authorrunning{J.  Krti\v{c}ka et al.}

\offprints{J. Krti\v{c}ka, \email{krticka@physics.muni.cz}}

\institute{Department of Physics and Astronomy, University of Glasgow,
           Glasgow G12 8QQ, UK
           \and
           \'Ustav teoretick\'e fyziky a astrofyziky P\v{r}F MU,
            CZ-611 37 Brno, Czech Republic
           \and
                   Astronomick\'y \'ustav, Akademie v\v{e}d \v{C}esk\'e
           republiky, CZ-251 65 Ond\v{r}ejov, Czech Republic
           \and
           Bartol Research Institute, University of Delaware, Newark,
           DE 19716, USA}

\date{Draft version \today}

\abstract{
The Discrete Absorption Components (DACs) commonly observed in the ultraviolet 
lines of hot stars have previously been modelled by dynamical simulations of 
Corotating  Interaction Regions (CIRs) in their line-driven stellar winds.
Here we apply the kinematic DAC inversion method of Brown et al. to 
the hydrodynamical CIR models
and test the reliability of the results obtained. We conclude that the inversion
method is able to recover valuable information on the velocity structure 
of the mean wind and to trace movement of velocity plateaux in
the hydrodynamical data, though the recovered density profile of the stream is correct
only very near to the stellar surface.
}

\maketitle

\section{Introduction}

The time-averaged structure of line-driven winds of hot OB stars can be
adequately described by time-independent wind models (eg.  Hillier \&
Miller \cite{hilmi}, Vink et al.
\cite{vikola}, Pauldrach et al. \cite{pahole}). However, winds of hot stars are
well-known for the variability manifested in different domains of the
electromagnetic spectrum. One of the most striking features in the spectra of
hot star winds is the presence of Discrete Absorption Components (DACs). These
DACs are superimposed on the (unsaturated) absorption part of P Cygni profiles
and move from the lower blue-shift to higher blue-shift  part of P Cygni
profiles (e.g. Prinja \& Howarth \cite{prihow}, Massa et al. \cite{mafuni},
Kaper et al. \cite{kaheful}, Prinja et al. \cite{primaful}).

The main DAC features, i.e. their recurrence and slow acceleration, can be
explained by the radiatively driven
hydrodynamical simulations of Cranmer and Owocki (\cite{crow}).
They assumed that DACs arise from Corotating Interaction Regions 
(CIRs), as first suggested by Mullan (\cite{mul}). The origin of CIR formation is not
yet known, though wind modulations by stellar pulsations or magnetic fields 
are often cited as possible drivers (Kaper et al. \cite{kaheful}, de Jong et
al. \cite{dejhenka}).
Thus it is worthwhile to study the DAC phenomenon from data diagnostically, 
independent of any underlying
driver model. A first step in this direction was made by Brown et al.
(\cite{brobaros}), who developed a kinematic data inversion method for the
inference of absorbing stream properties from data on recurrent DACs.
The optical depth variations calculated using the hydrodynamical
structures of the Cranmer and Owocki (\cite{crow}) models on the other hand
arise mainly  due to the azimuthally localized
velocity plateau (so-called Abbott kinks) propagating
upstream in the stellar wind. These  velocity plateaux are caused by the
presence of radiative-acoustic waves (the so-called Abbott waves,
Abbott~\cite{abb}) moving with velocity higher than the sound speed.  
Thus, it is interesting to study how well the kinematic inversion method for
line-profiles performs using the profiles from hydrodynamical CIR model 
calculations, specifically to see what the method recovers in terms of density and velocity
variations.

\section{Kinematic inversion formulation}
\label{kininfor}

The basic method of the DAC data inversion of Brown et~al.~(\cite{brobaros})
can be summarised as follows. They assume that DACs arise as the result
of an azimuthally varying mass loss rate at the stellar surface, giving
a density pattern in the wind that also varies with azimuth, and is
then transported through the wind along the streamlines of an
\emph{axisymmetric} velocity law~$v_r=v_r(r)$. The assumed axisymmetry
of the wind velocity has several consequences:
\begin{enumerate}
\item From mass continuity it follows that the radial mass flux $nv_r$
falls off with radius along any streamline as~$1/r^2$, which is the
purely geometrical factor expected for essentially 
spherical
wind expansion
(which we view only in the equatorial plane~-- see below).
\label{itone}
\item From~\ref{itone}, it further follows that the density itself at any point
along a streamline is related to the mass flux $n_0v_0$ at the point on 
the
surface from which that streamline emanates by
\begin{equation}
n(r)=\frac{v_0R^2}{v_rr^2}n_0 \quad \hbox{(along a streamline),}
\label{n-var-stream}
\end{equation}
where $R$ is the stellar radius.
\item The streamlines from any two points on the surface (at $\phi_1$
and~$\phi_2$, say) will always have identical shapes for an axisymmetric
velocity law, but will be shifted relatively in azimuth by $\phi_2-\phi_1$ 
at
all radii. This means that the density pattern introduced into the wind
at the surface will be preserved at all radii, but will be
shifted in $\phi$ by the change in azimuth $\Delta\phi(r)$ of the
streamlines due to rotation. In the corotating frame this is given by
\begin{equation}
\Delta\phi(r)=\int_R^r\left[\Omega_0-\Omega(r)\right]
\frac{\mathrm{d}r}{v_r(r)},
\end{equation}
where $\Omega(r)r=v_\phi$ is the rotational speed in the static frame
and $\Omega_0=\Omega(R)$ is the angular velocity of the surface.
(The density will also be scaled according to Eq.~(\ref{n-var-stream})
along the streamlines.)
\end{enumerate}
Putting these points together we can relate the physical density
$n(r,\phi)$ of the wind at any point to the surface density
profile~$n_0(\phi) = n(R,\phi)$ by
\begin{equation}
n(r,\phi) = \frac{v_0R^2}{v_rr^2}n_0[\phi-\Delta\phi(r)],
\label{n-variation}
\end{equation}
so that the optical depth variation at the velocity~$v=v_r(r)$ 
corresponding
to this radius is (from $\tau = n/v_r^\prime$, in suitable units)
\begin{equation}
\tau(v_r,\phi)\equiv n(r,\phi)/{v_r}^\prime
                 = {\cal P}(v_r)\,n_0(\phi-\Delta\phi(r)),
\label{nstream}
\end{equation}
where
\begin{equation}
       {\cal P}(v_r) = \frac{v_0R^2}{v_r{v_r}^\prime r^2}
\label{Pdef}
\end{equation}
gives the overall ($\phi$-independent) weighting of the optical depth
at each velocity and $v_r^\prime$ is the velocity gradient.

The inversion method of Brown et~al.~(2003) uses the variation of the
observed optical depth (using the Sobolev approximation 
Eq.(\ref{nstream}))
with time (i.e.,~with observer azimuth, $\phi=\Omega_0 t$) to determine
the optical depth weighting function~${\cal P}(v_r) $ and the spiral
law $\Delta\phi(r)$, which then give the wind velocity law $v_r(r)$ and
rotation law~$\Omega(r)$. Finally, from the known optical depth variations
with azimuth at fixed radius (i.e.,~velocity) it is possible to
calculate the surface density profile~$n_0(\phi_0)$.

When the observed dynamical spectrum is consistent with the assumption
of an axisymmetric velocity law it is easy to determine ${\cal P}(v)$
and $\Delta\phi(r)$. Integration of Eq.~(\ref{nstream}) over~$\phi$
at fixed~$v_r$ gives
\begin{equation}
        {\cal P}(v) \propto \int_0^{2\pi} \tau(v,\phi)\,\mathrm{d}\phi.
\label{Pcalc}
\end{equation}
Alternatively, if it is possible to identify the path of a single matter
stream in the $(v_r,\phi)$-plane, such as the peak of the density pattern
in azimuth at each~$v_r$, the optical depth variation along this matter
stream can be used to define~${\cal P}$, although this is likely to be
more sensitive to data errors on the dynamical spectrum.
To obtain $\Delta\phi(r)$ it is necessary to identify just such a matter
stream, and to define its path in the $(v_r,\phi)$ plane.

\section{Dynamical DAC prediction and kinematic model inversions}

\subsection{DAC prediction}

For the hydrodynamical prediction of DACs we apply a similar method to Cranmer 
\& Owocki (\cite{crow}). We assume that DAC features in the spectra of OB stars 
are caused by CIRs in the stellar wind induced by a hot spot on the stellar
surface. Although CIR formation in a radiatively
driven wind  is inherently three-dimensional, to make the problem more
tractable we confine the hydrodynamical calculations only to the equatorial
plane, $\theta=\pi/2$.

In order to obtain static DAC models we relax the time-dependent hydrodynamical
equations to a steady-state.
For the solution of time-dependent hydrodynamical equations we use the code VH-1 
developed by J. M. Blondin and collaborators. The equations solved are equation
of continuity (conservation of mass) in the form
\begin{equation}
\frac{\partial\rho}{\partial t}+
\frac{1}{r^2}\frac{\partial}{\partial r}(\rho v_r r^2)+
\frac{1}{r\sin\theta}\frac{\partial}{\partial \phi}(\rho v_\phi)=0,
\end{equation}
and the conservation of radial ($r$) and azimuthal ($\phi$)
components of momentum,
\begin{gather}
\frac{\partial v_r}{\partial t}+v_r\frac{\partial v_r}{\partial r}+
\frac{v_\phi}{r\sin\theta}\frac{\partial v_r}{\partial\phi}=
\frac{v_\phi^2}{r}-\frac{1}{\rho}\frac{\partial P}{\partial
r}+g_{\mathrm{eff}}+g^{\mathrm{lines}},\\
\frac{\partial v_\phi}{\partial t}+v_r\frac{\partial  v_\phi}{\partial r}+
\frac{v_\phi}{r\sin\theta}\frac{\partial v_\phi}{\partial\phi}=
-\frac{v_rv_\phi}{r}-\frac{1}{\rho r\sin\theta}\frac{\partial P}{\partial\phi},
\end{gather}
where $\rho$ is the local wind density, $v_r$ and $v_\phi$ are the radial and azimuthal
components of wind velocity, $t$ is time, and gas pressure $P$ is evaluated
using the perfect gas equation of state.

The effective gravitational acceleration is given by
\begin{equation}
g_\mathrm{eff}=-\frac{GM(1-\Gamma)}{r^2},
\end{equation}
where $G$ is the gravitational constant, $M$ is the stellar mass and the 
Eddington factor
$\Gamma$ accounts for the radiative force due to electron scattering. The
radiative acceleration due to lines in the Castor, Abbott \& Klein (\cite{cak}) approximation  is
given by
\begin{equation}
g^\mathrm{lines}=kf A(\psi) \frac{GM\Gamma}{r^2}
\left(\frac{n_\mathrm{e}/W}{10^{11}\mathrm{cm}^{-3}}\right)^\delta
\left(\frac{1}{\kappa_\mathrm{e}v_\mathrm{th}\rho} 
\left|\frac{\partial v_r}{\partial r}\right|\right)^\alpha,
\end{equation}
where $k$, $\alpha$ and $\delta$ are line-force parameters (Abbott
\cite{abpar}, Puls et al. \cite{silstat}), $f$ is the finite-disk correction
factor (Friend \& Abbott \cite{fa}, Pauldrach et al.~\cite{ppk}), $n_\mathrm{e}$
is the electron number density, $W$ is the dilution factor,
$\kappa_\mathrm{e}$ is the opacity due to the
electron scattering,
and $v_\mathrm{th}$ is a 
fiducial
ion thermal speed (by definition
hydrogenic). The function $A(\psi)$ accounts for the force enhancement due to 
the spot, where 
the
corotating azimuthal angle $\psi$ is given by
\begin{equation}
\psi=\phi-\Omega_0 t,
\end{equation}
with the star's rotational angular velocity $\Omega_0=V_\mathrm{rot}/R_*$.
The force amplitude has the form
\begin{equation}
A(\psi)=1+A_0 \exp\left[-\left(\psi-\psi_0\right)^2/\sigma^2\right],
\end{equation}
where $A_0$ is the dimensionless spot amplitude (corresponding to the flux
enhancement), $\psi_0$ is the corotating azimuthal position of the center of the
spot and $\sigma$ describes the spatial extent of the spot on the star. Thus,
contrary to the paper of Cranmer \& Owocki (\cite{crow}), we have neglected
limb-darkening.

The flow variables are specified on a fixed two-dimensional mesh in radius and
azimuth. We use 250 radial zones, extending from $r=R_*$ to $r=11R_*$ with
higher concentration of zones near the stellar base. The spatial zone width
radially increases by $1\%$ per zone. The azimuthal mesh contains 180 constantly spaced
zones, ranging from $0$ to $2\pi$.  The boundary conditions are specified in the
same way as in  Owocki et al. (\cite{ocb}) and Cranmer \& Owocki (\cite{crow}).
As an initial condition we use the relaxed one-dimensional model of 
Krti\v cka \& Kub\'at (\cite{kkii}).

The DAC profile is calculated using the solution of the radiative transfer equation
in the absence of any source function term in a point-star approximation as
\begin{equation}
\label{taurov}
\tau(v,\phi)=\int_{R_*}^{\infty}\kappa\rho(r,\phi)
\exp\left[-\left(v_r(r,\phi)-v\right)^2/v_\mathrm{G}^2\right]\mathrm{d}r, 
\end{equation}
where 
$\kappa$ is line-opacity (its arbitrary
value is selected to obtain optical depths of order unity) and $v_\mathrm{G}$
descibes profile broadening as a sum of temperature broadening and
microturbulent broadening by
\begin{equation}
v_\mathrm{G}^2=v_\mathrm{th}^2+v_\mathrm{turb}^2,
\end{equation}
where $v_\mathrm{turb}$ is a microturbulent velocity.

Surprisingly, the moving absorption features (DACs) in the spectra
calculated from the hydrodynamical simulations are not predominantly caused by
density variations. This can be seen from the fact that the wind optical depth can be approximated by the so-called
Sobolev approximation in the form of
\begin{equation}
\label{sobap}
\tau\sim\frac{\rho}{v_r^\prime }.
\end{equation}
Even with relatively small variations of wind density and velocity the
variations of velocity gradient and also of the Sobolev optical depth may be
large. This is especially the case when a typical feature of our hydrodynamical
calculations develops, the so-called Abbott kink. In the region of an Abbott kink
the velocity derivative changes its sign from positive to negative and back.
Thus, as can be seen from Eq.(\ref{sobap}), the optical depth variations due to
the Abbott kink may be huge.

\subsection{Kinematic model inversion}

For the inversion of the DAC data (calculated using hydrodynamical simulations)
we use essentially the same code as Brown et al. (\cite{brobaros}). The DAC data
velocities are normalised with respect to the maximal velocity obtained in
the hydrodynamical simulations, and only optical depth variation data
for lower velocities are used in the kinematic model inversion procedure.
Finally, for the inversion of the velocity law, we take into account that the
DAC profiles are calculated using data from only finite intervals of radius.

\section{Results}

\begin{figure*}
\resizebox{0.5\hsize}{!}{\includegraphics{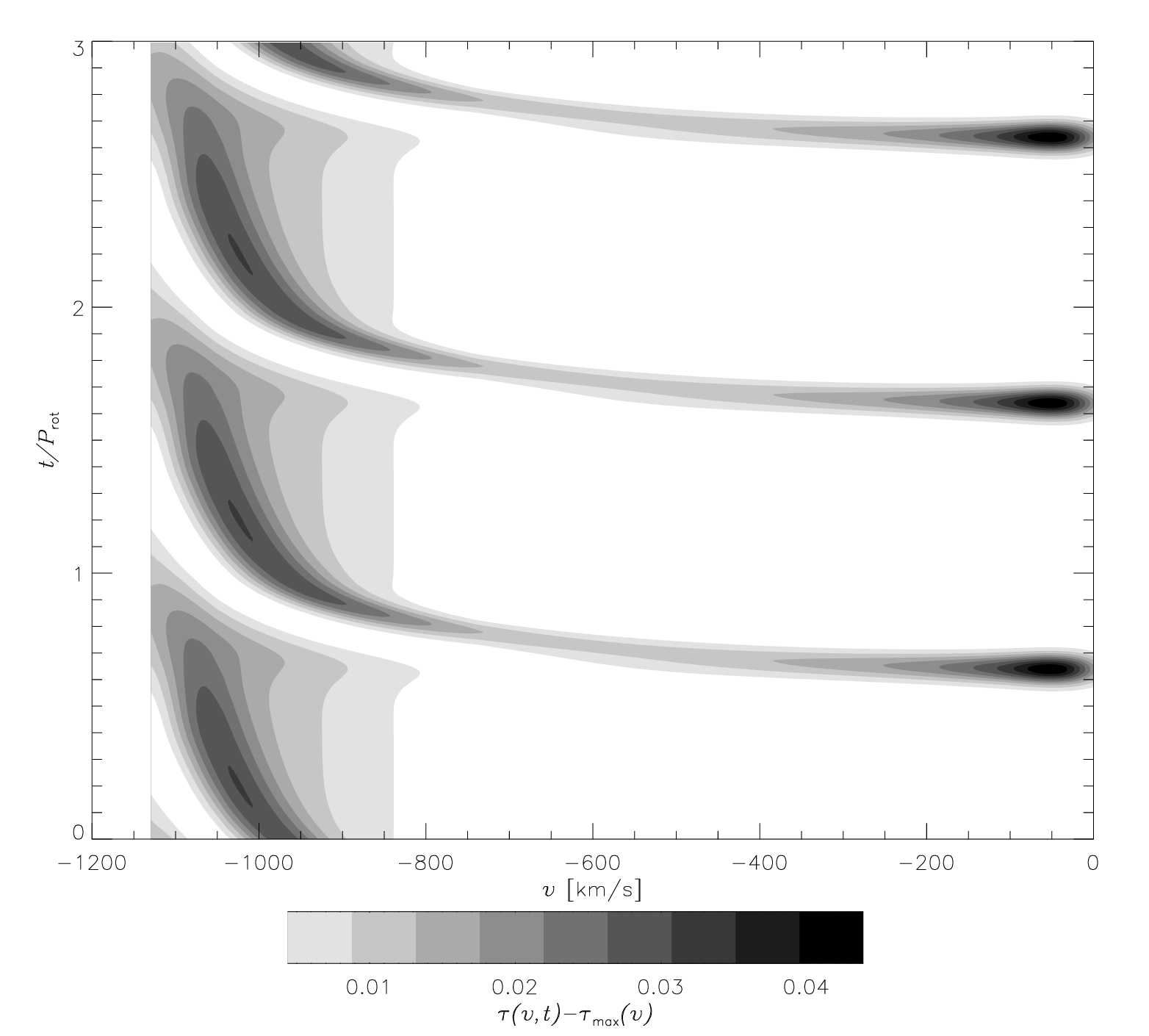}}
\resizebox{0.5\hsize}{!}{\includegraphics{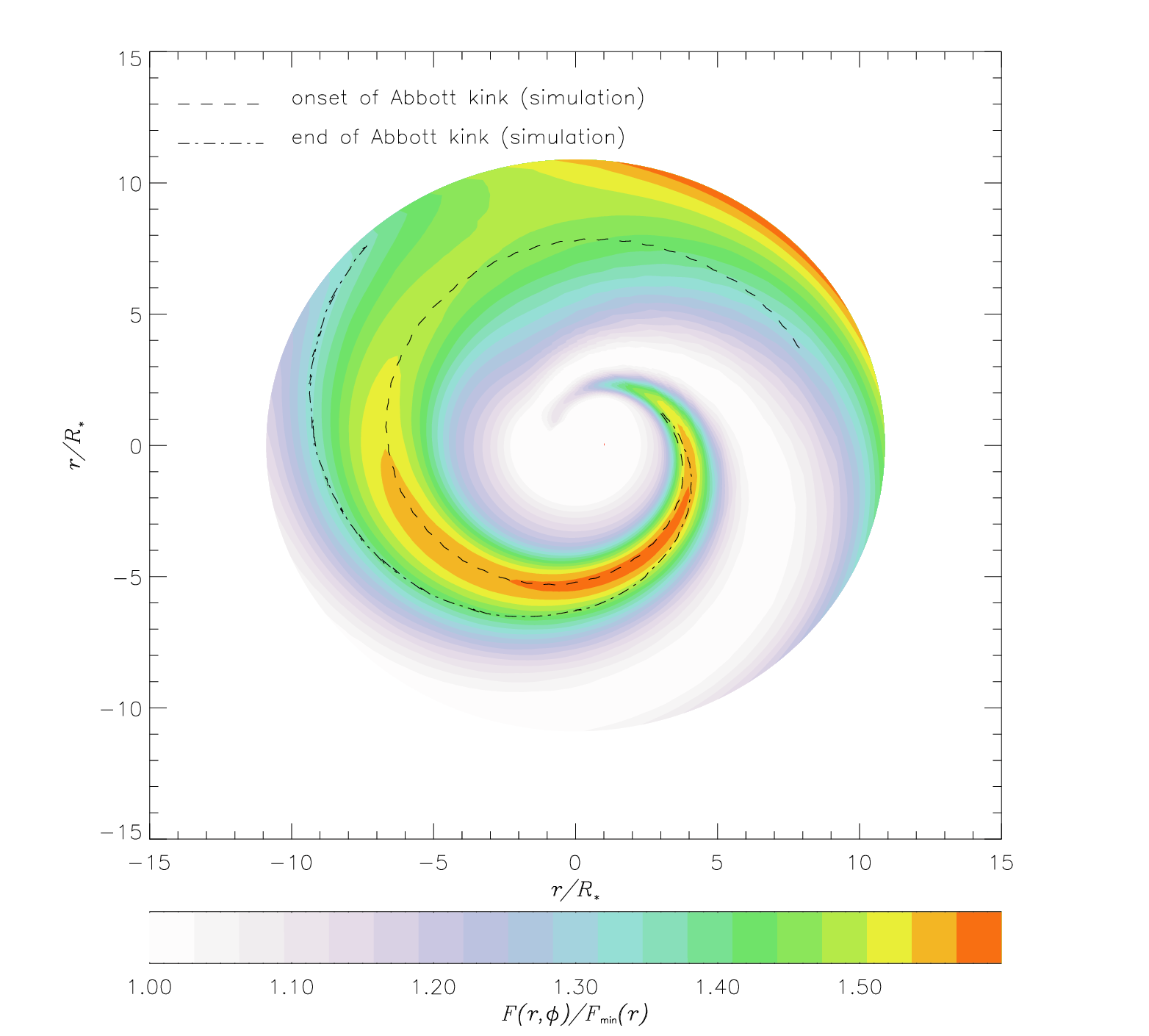}} \vskip -2mm
\resizebox{0.5\hsize}{!}{\includegraphics{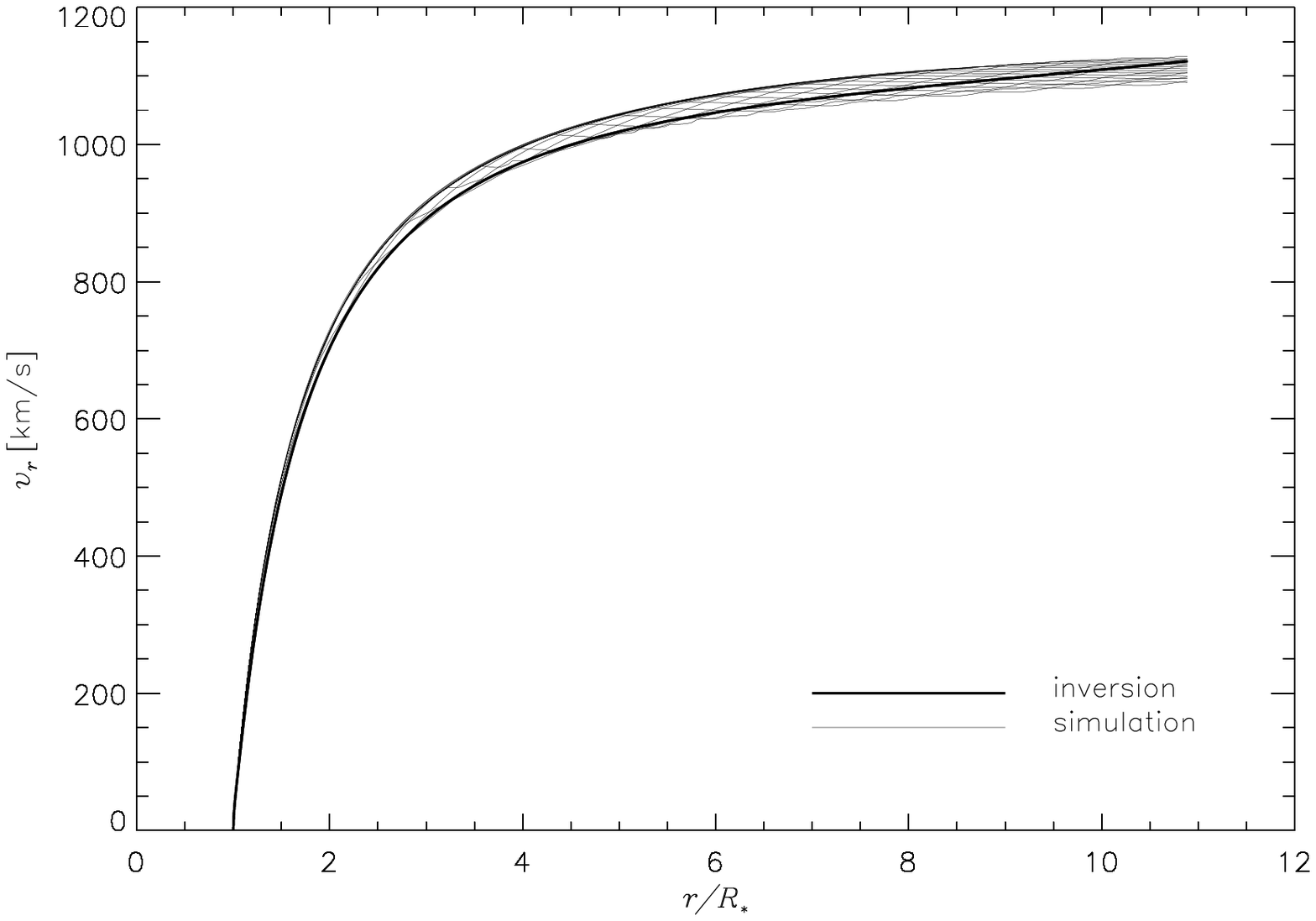}} 
\resizebox{0.5\hsize}{!}{\includegraphics{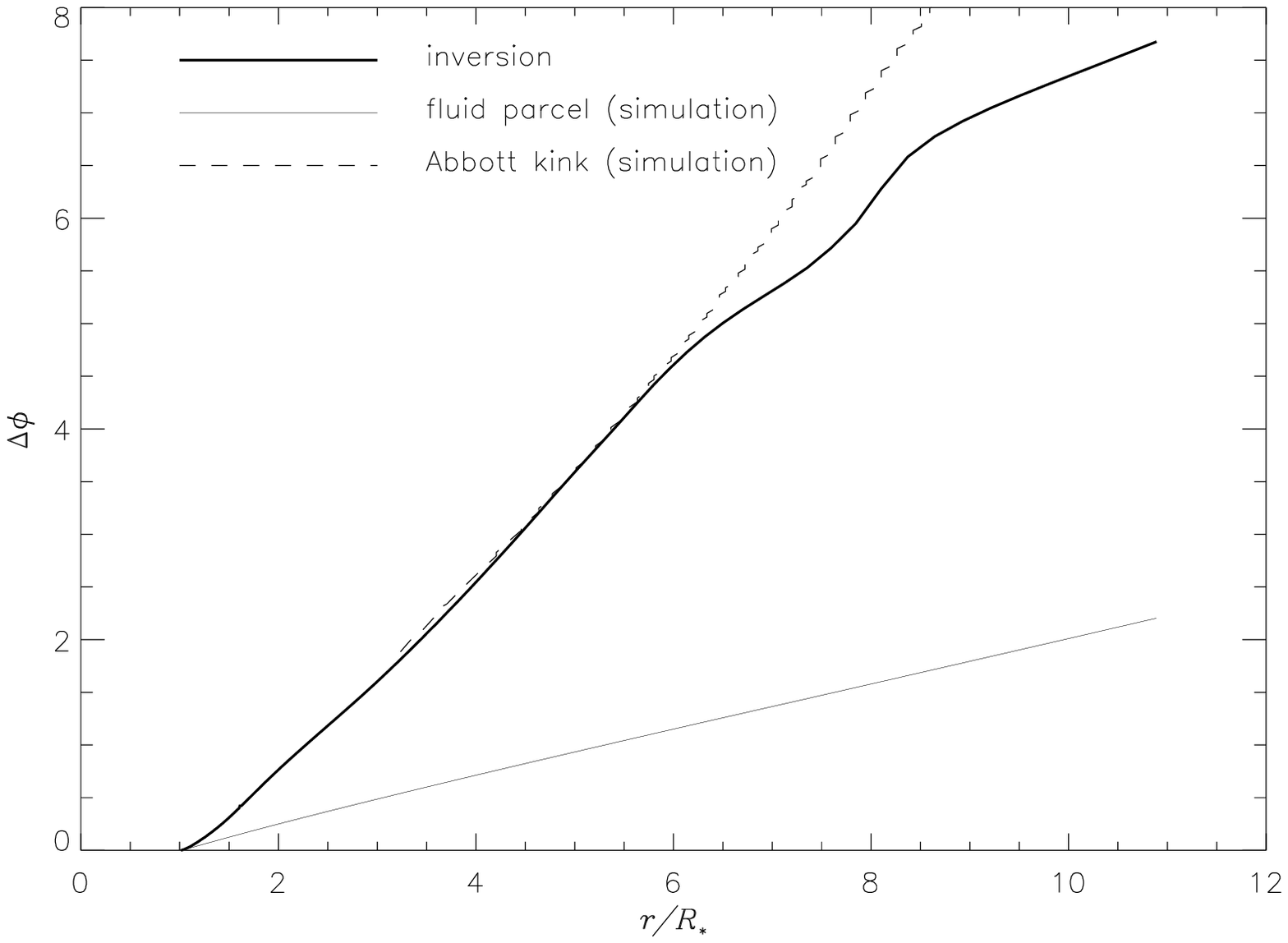}}\vskip -2mm
\resizebox{0.5\hsize}{!}{\includegraphics{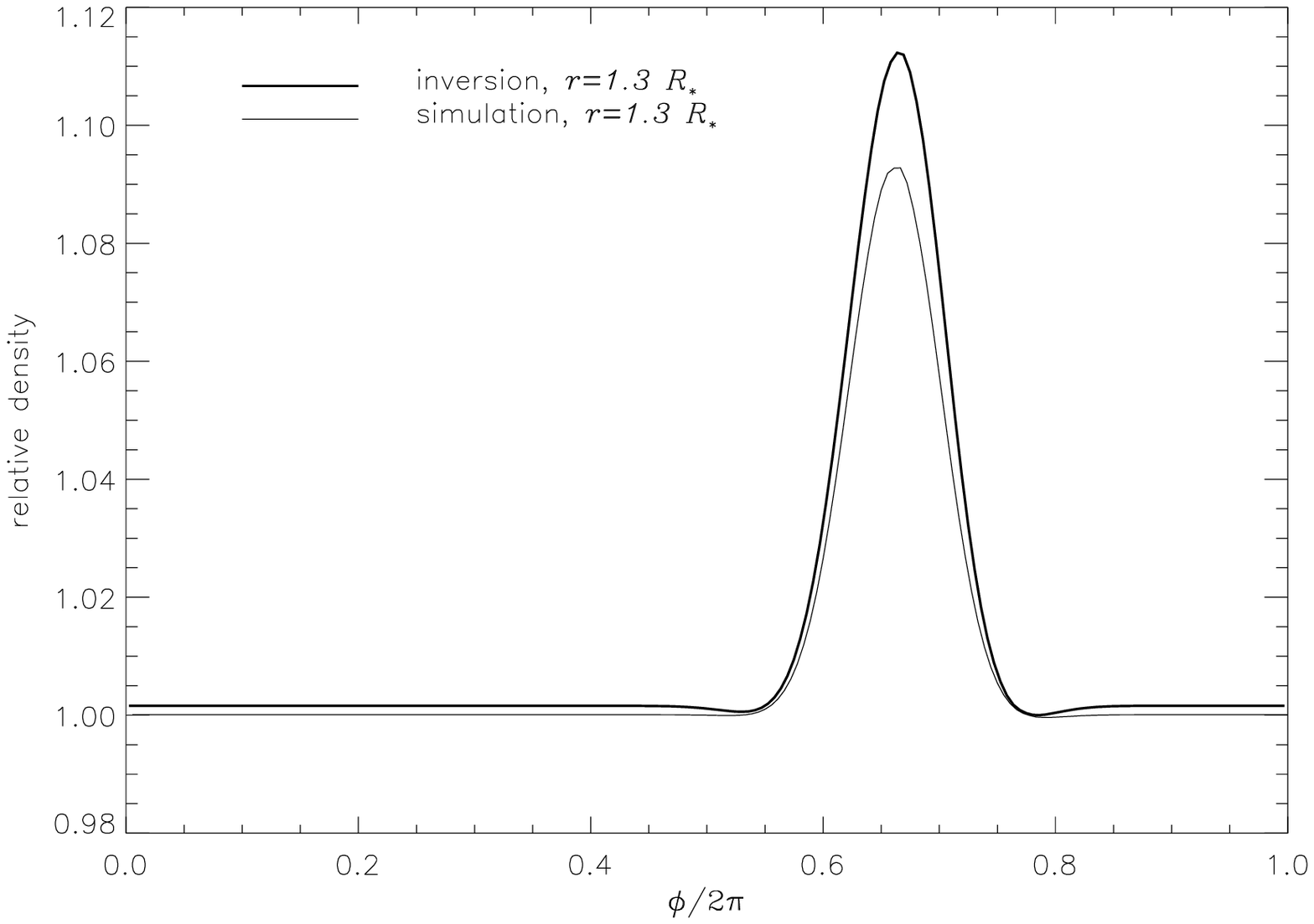}}
\resizebox{0.5\hsize}{!}{\includegraphics{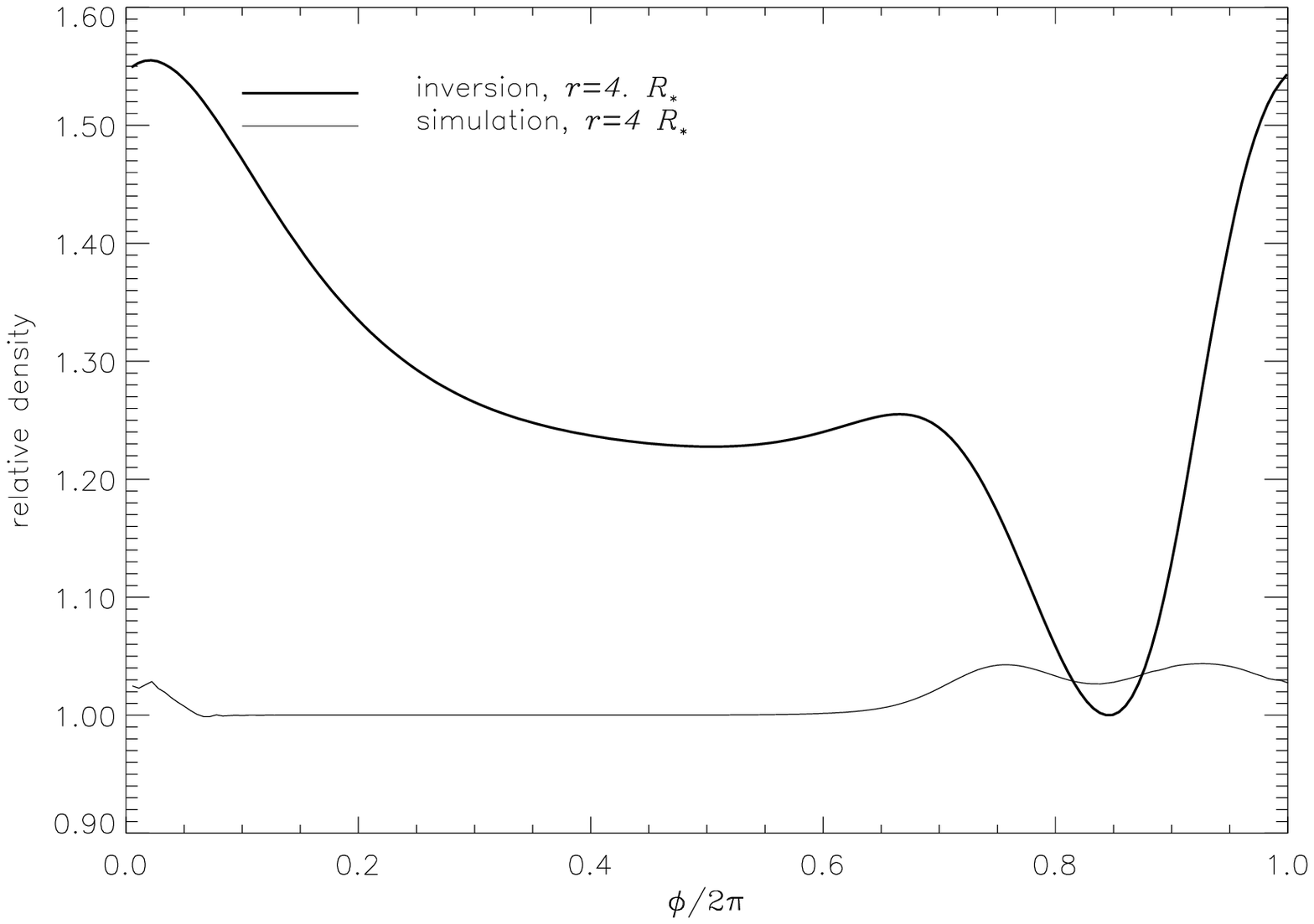}}\vskip -2mm
\caption[]{Comparison of hydrodynamical simulations and inverted data
for
a
spot with relative amplitude $A=0.05$.
{\em Upper left panel:}
Variation of DAC optical depth
(calculated as a difference to minimal absorption template)
derived using the data from simulations.
{\em Upper right panel:} Inverted variations of effective density and the
locations of 
the
onset and end of Abbott kink.
{\em Middle left panel:}
Radial velocity $v_r(r,\phi)$ taken from simulations at
18 equally spaced azimuths
and 
the
inverted mean radial velocity $v_r(r)$.
The Abbott kink in the hydrodynamical calculations can be clearly recognized.
{\em Middle right panel:}
Radial variation of azimuths $\Delta\phi$ of a fluid parcel 
relative to the spot, inverted location of density peak in azimuth and
the azimuth of the onset of Abbott kink.
%
{\em Lower panels:}
Variation of density from the simulations $\rho(r,\phi)$ relative to the the
density of 1D
flow without a spot $\rho_\mathrm{1D} (r)$ and
inverted density variation $F(r,\phi)$ for two different radii.
}
\label{spot05}
\end{figure*}

\begin{figure*}
\resizebox{0.5\hsize}{!}{\includegraphics{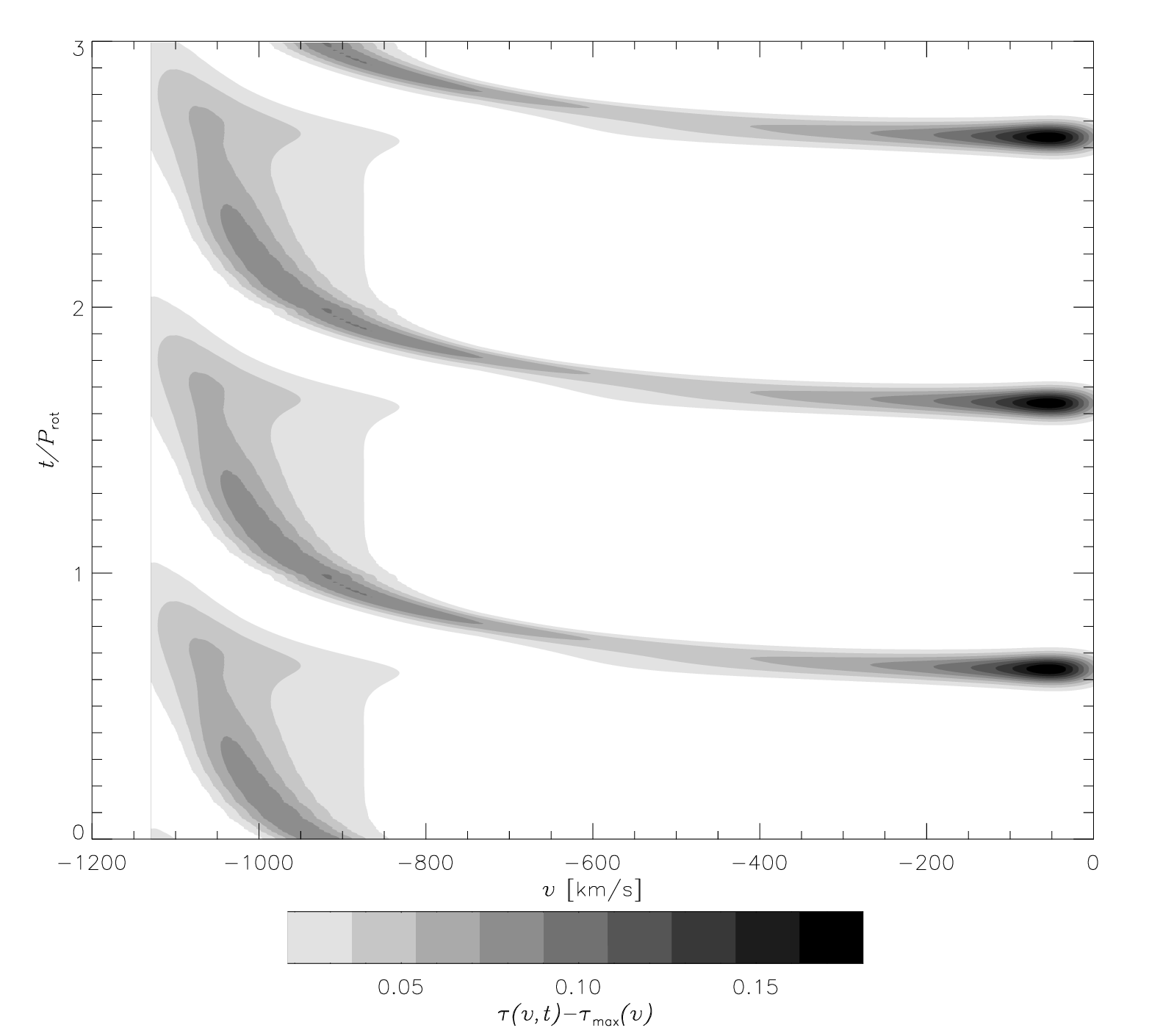}}
\resizebox{0.5\hsize}{!}{\includegraphics{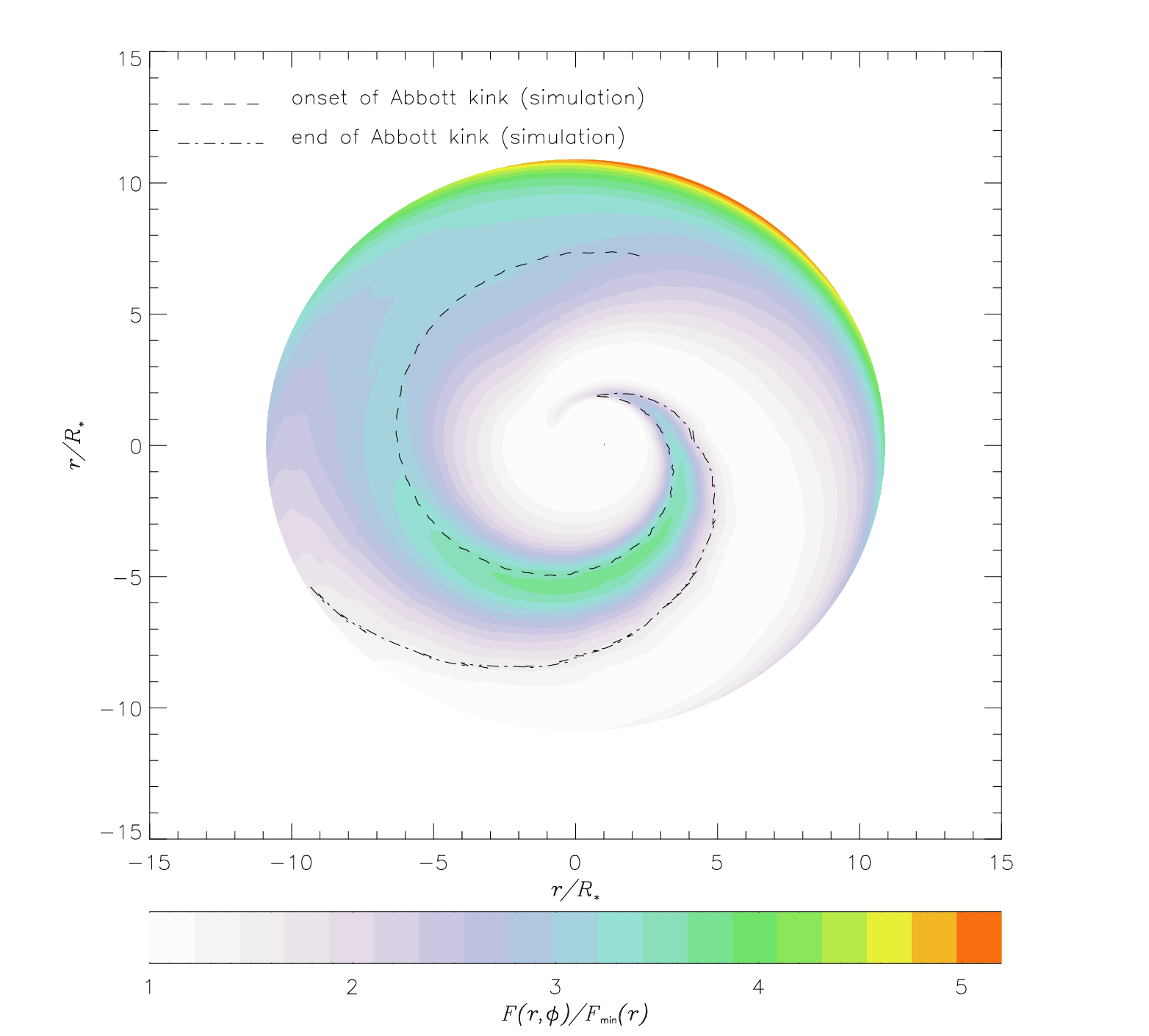}} \vskip -2mm
\resizebox{0.5\hsize}{!}{\includegraphics{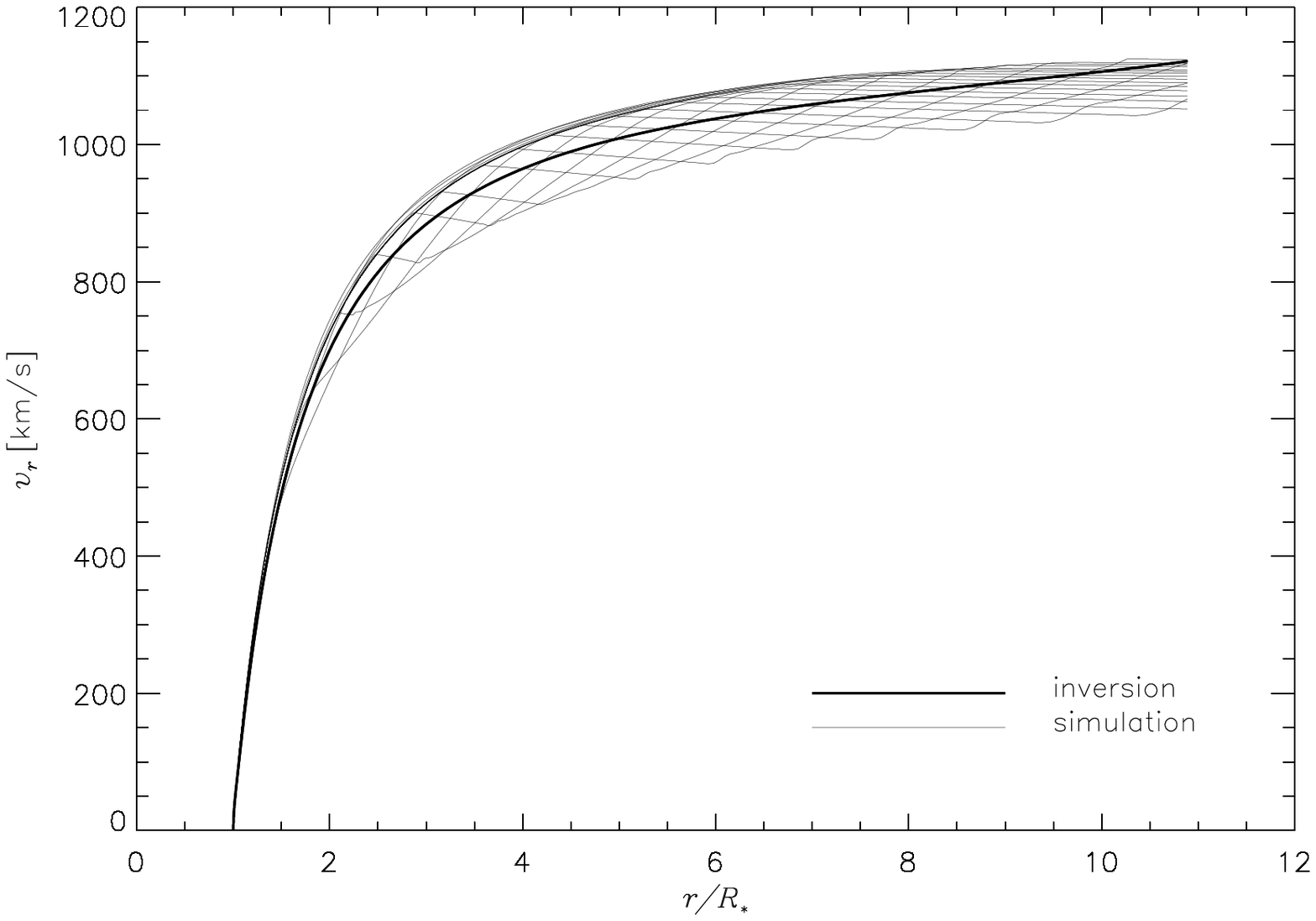}}
\resizebox{0.5\hsize}{!}{\includegraphics{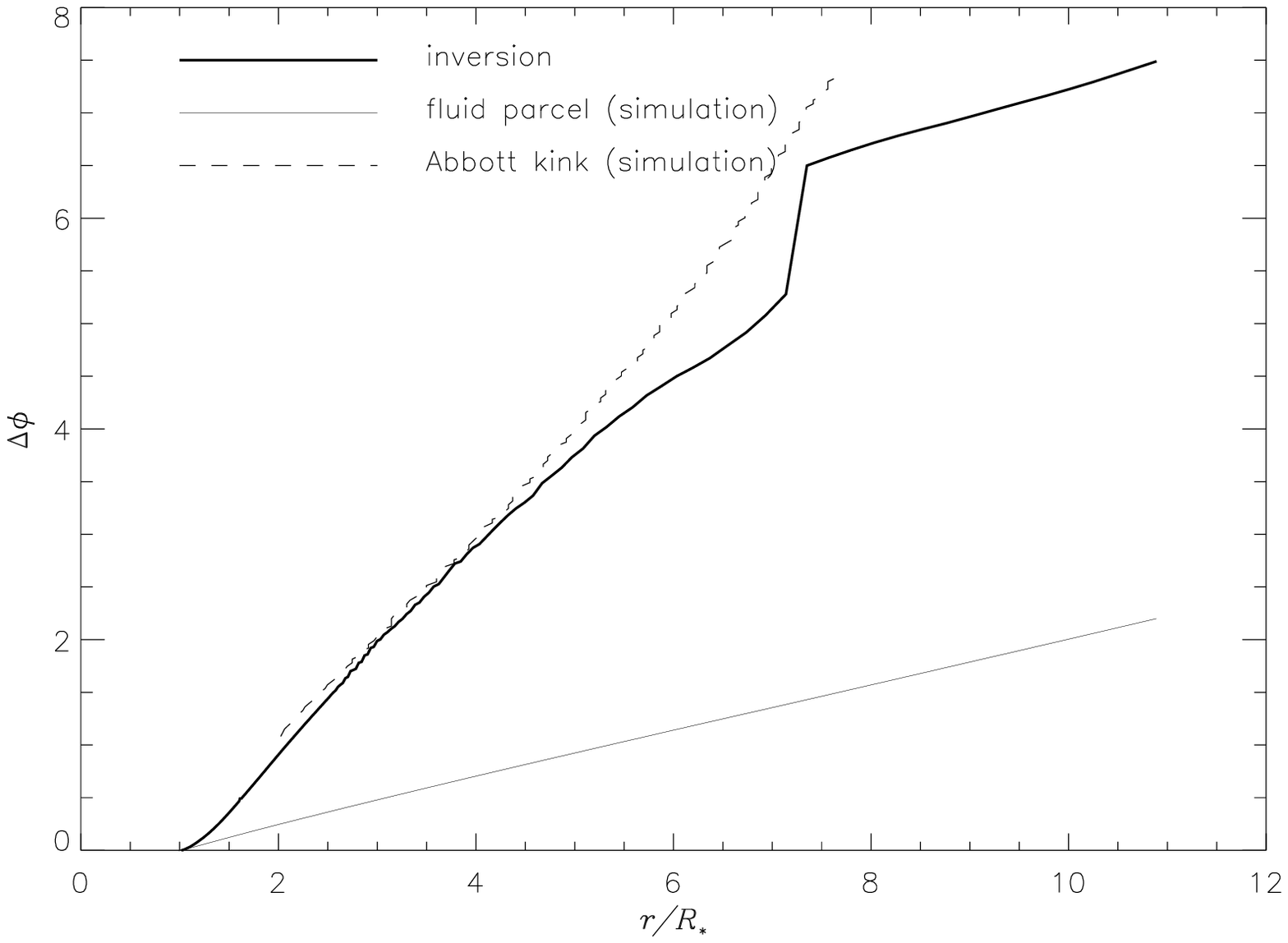}}\vskip -2mm
\resizebox{0.5\hsize}{!}{\includegraphics{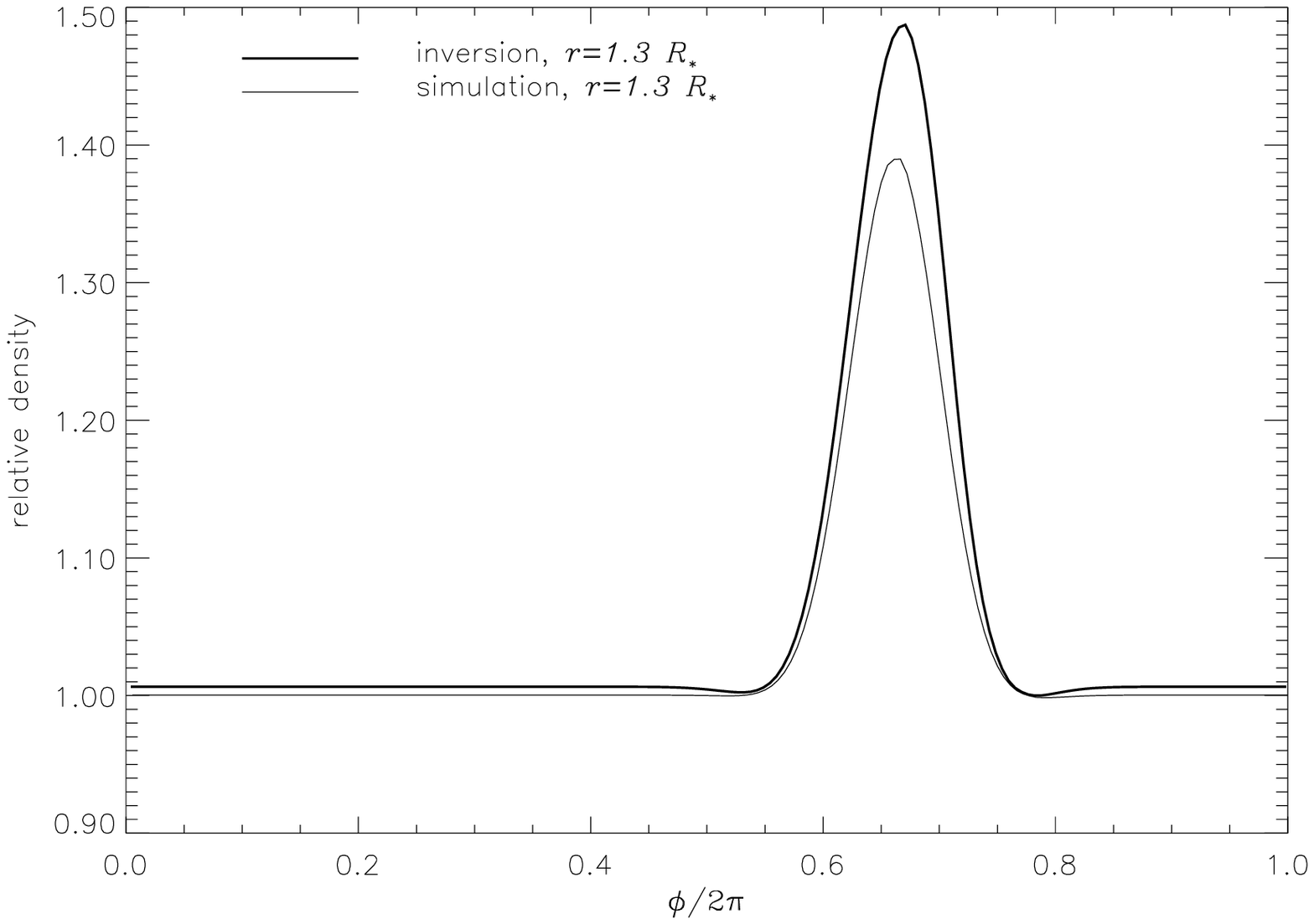}}
\resizebox{0.5\hsize}{!}{\includegraphics{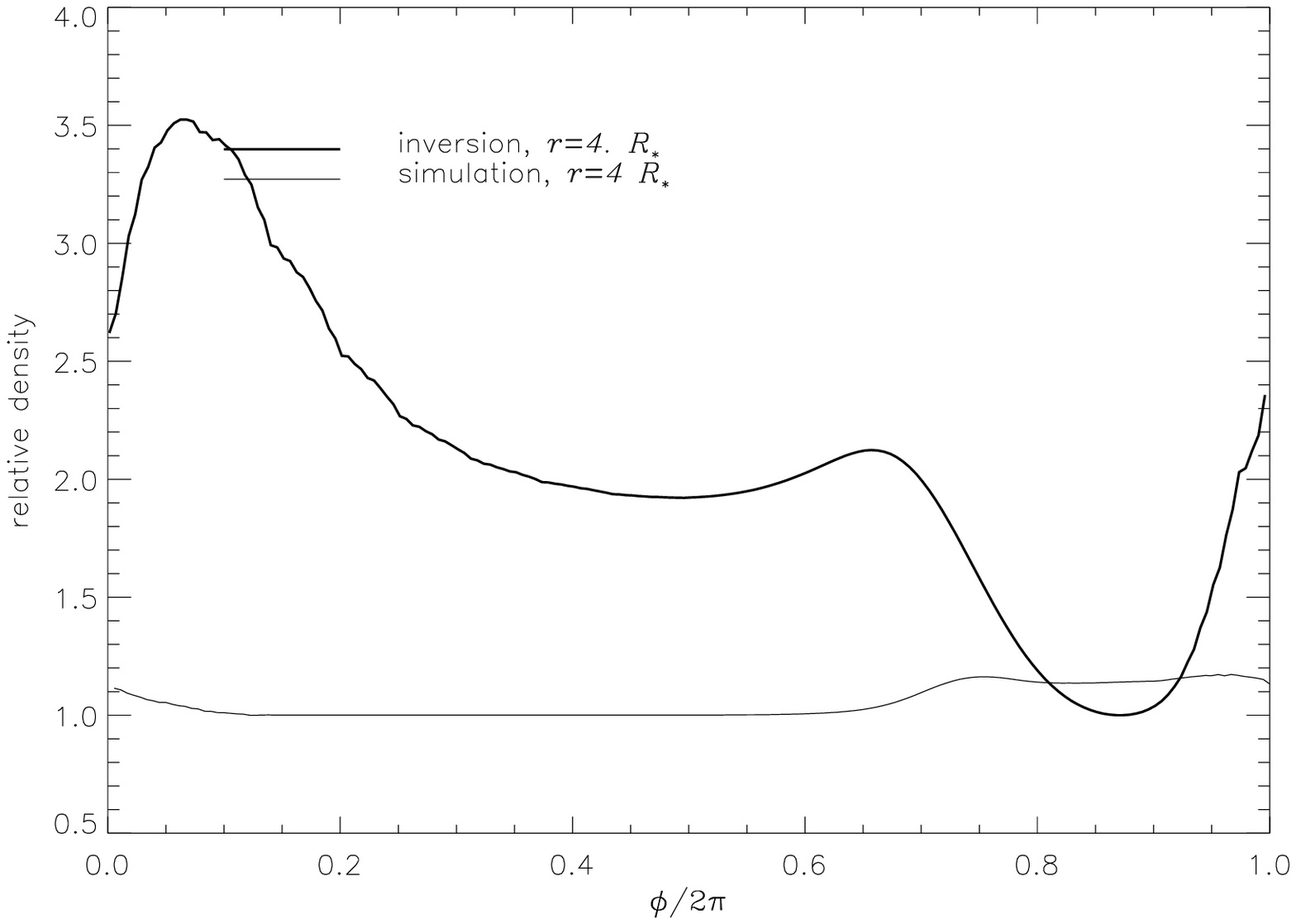}}\vskip -2mm
\caption[]{The same as Fig.\protect\ref{spot05}, however
for spot with relative amplitude $A=0.2$.}
\label{spot2}
\end{figure*}

\begin{figure*}
\resizebox{0.5\hsize}{!}{\includegraphics{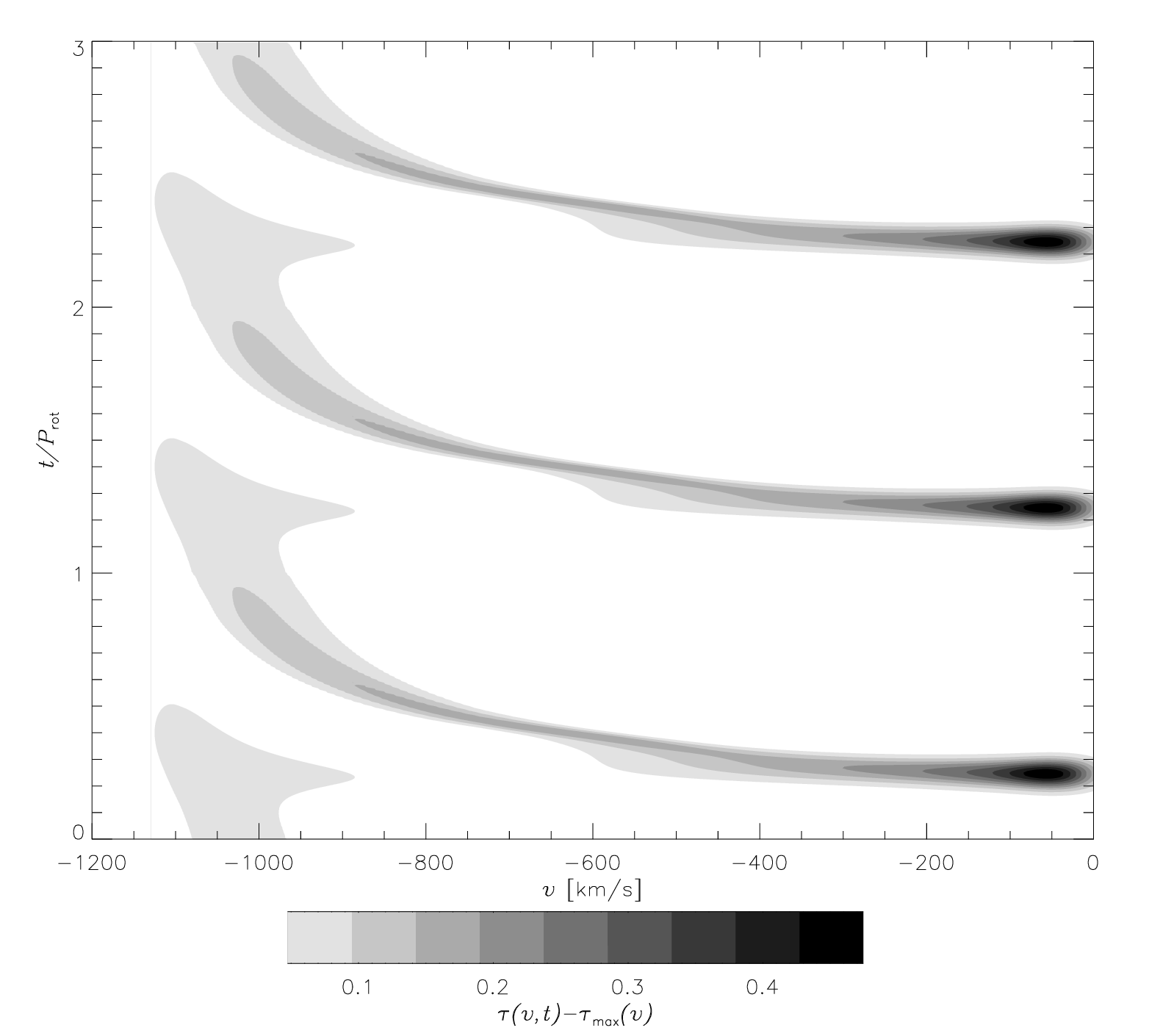}}
\resizebox{0.5\hsize}{!}{\includegraphics{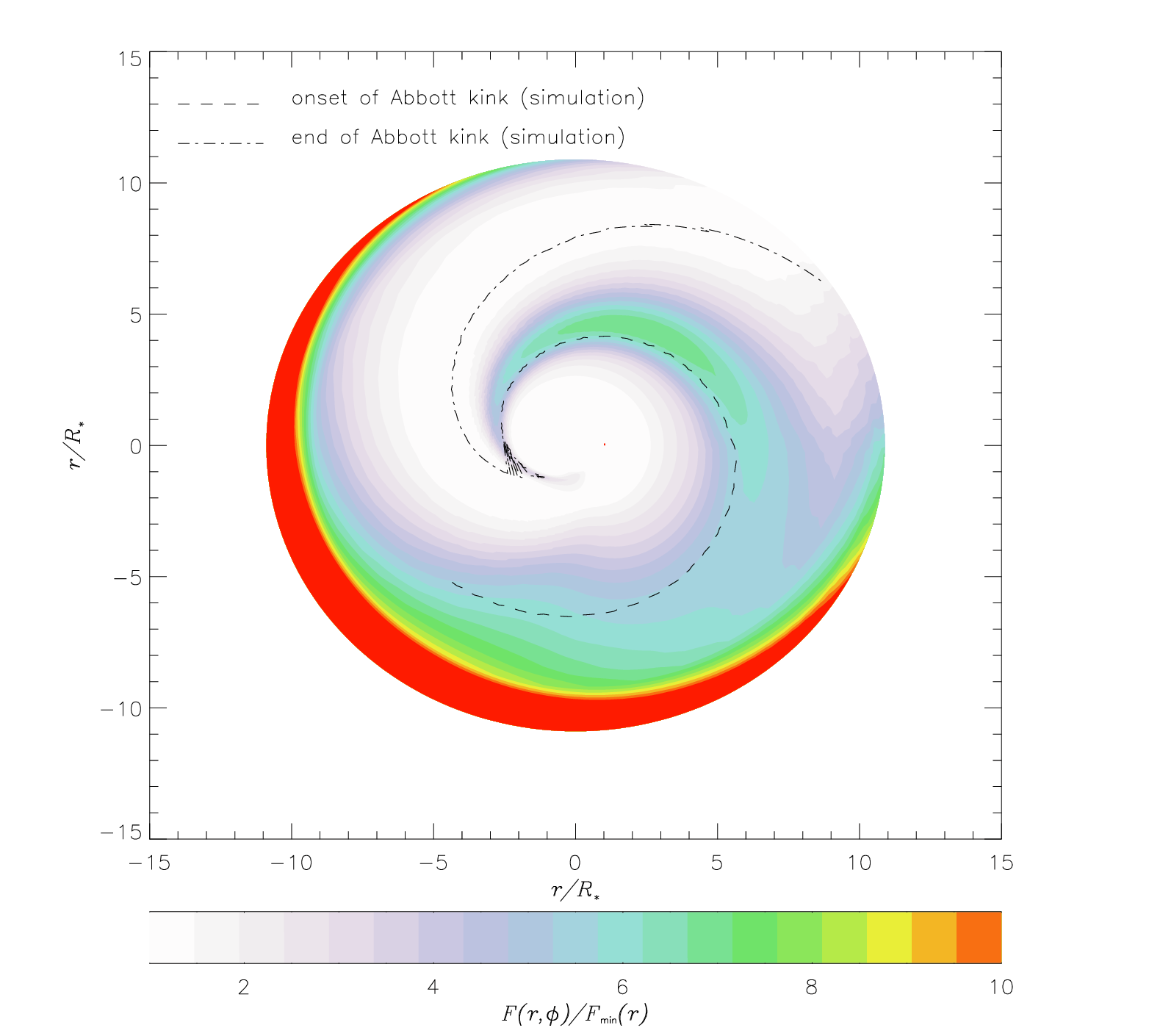}} \vskip -2mm
\resizebox{0.5\hsize}{!}{\includegraphics{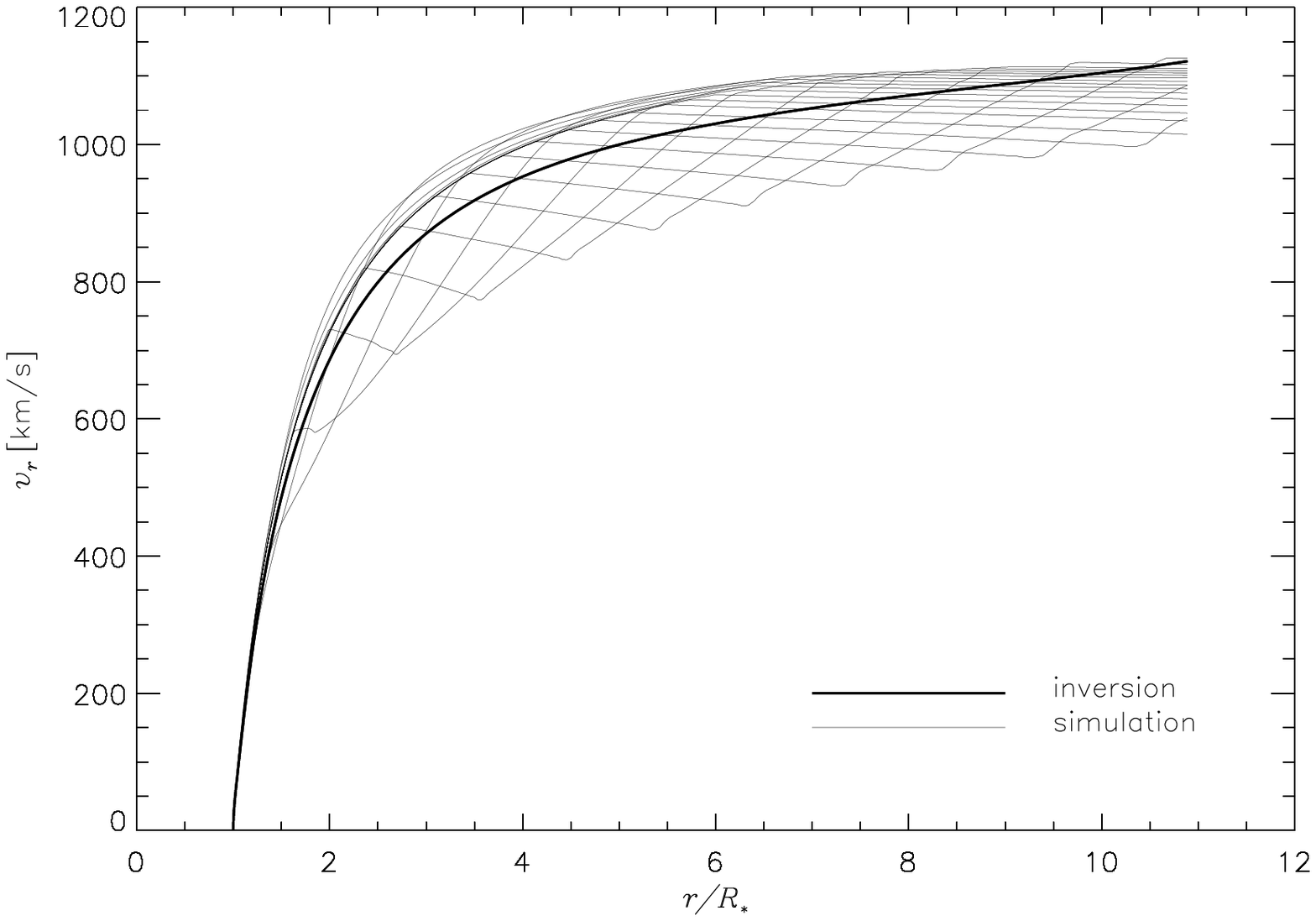}}
\resizebox{0.5\hsize}{!}{\includegraphics{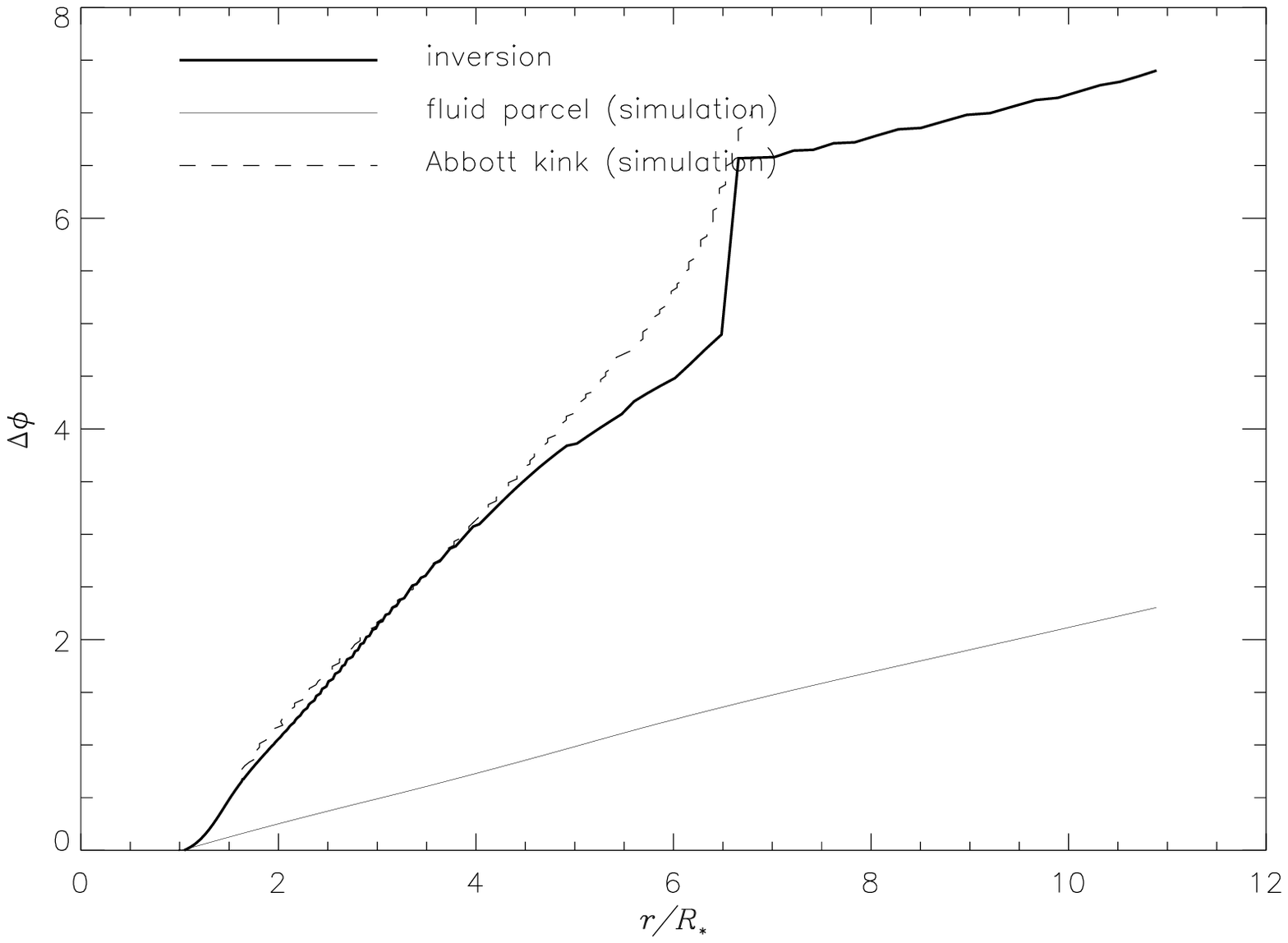}}\vskip -2mm
\resizebox{0.5\hsize}{!}{\includegraphics{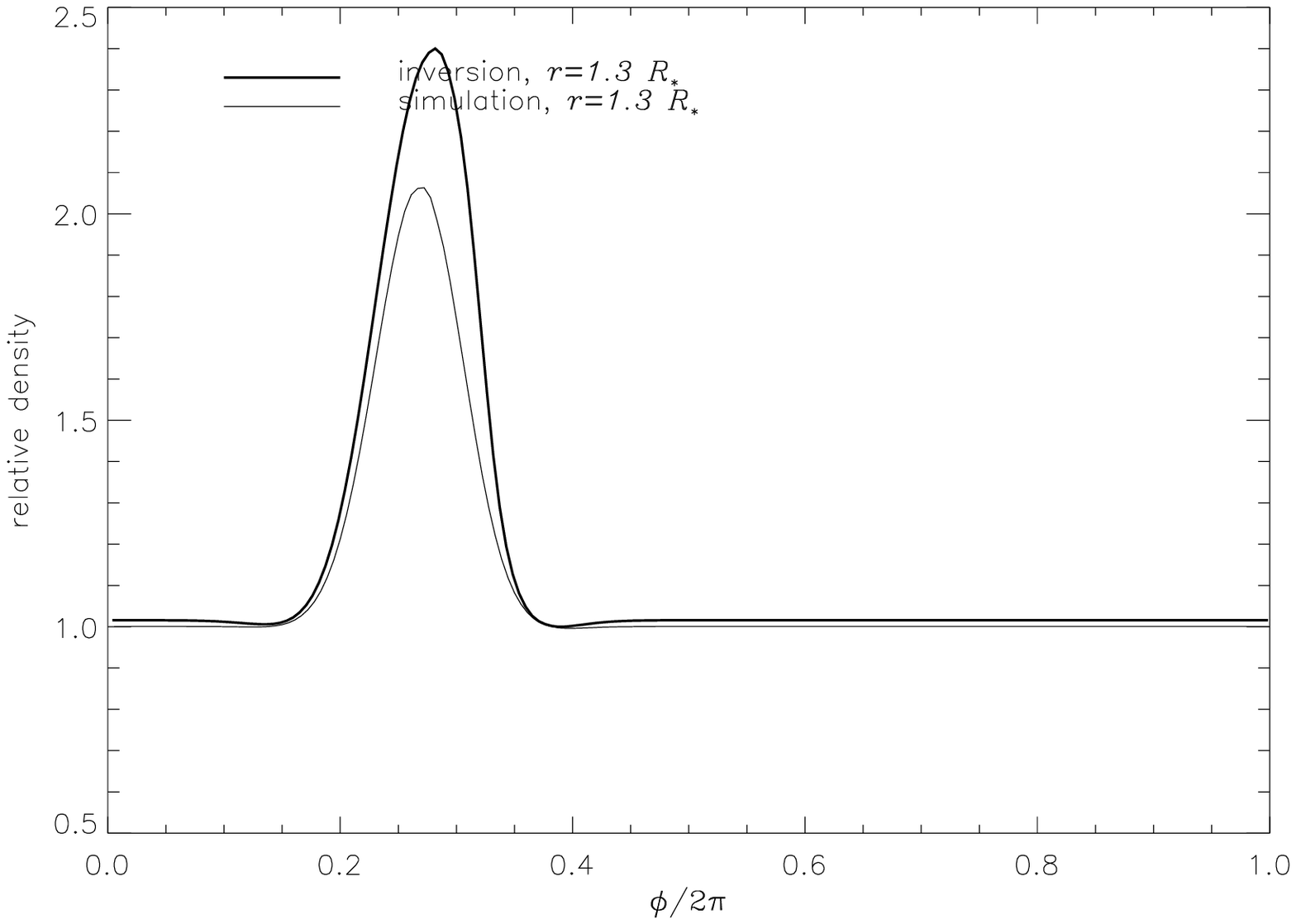}}
\resizebox{0.5\hsize}{!}{\includegraphics{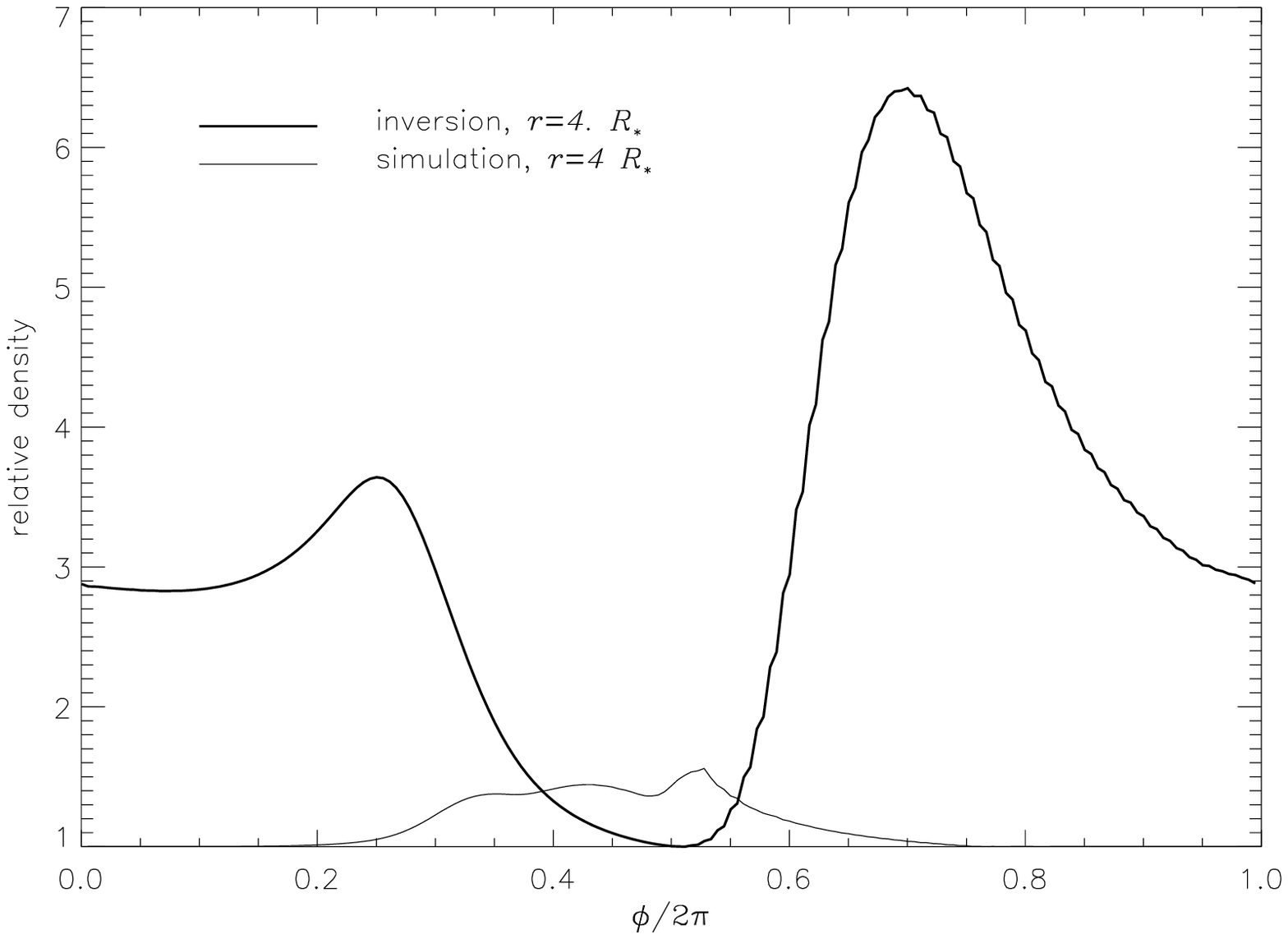}}\vskip -2mm
\caption[]{The same as Fig.\protect\ref{spot05}, however 
for spot with relative amplitude $A=0.5$.}
\label{spot5}
\end{figure*}

For our simulated study we selected the parameters of the star HD 64760 because
its wind variability was extensively
studied in the course of the IUE ``MEGA'' campaign (Prinja et al. \cite{megaprimaful}).
Stellar and wind parameters used for the hydrodynamical simulations are given in
Table~\ref{hdpar}. Stellar parameters were taken from Hoffleit \& Jaschek
(\cite{hofjas}),  Humphreys \& McElroy (\cite{humce}), and Schaller et al.
(\cite{schsch}, see Fullerton et al. \cite{fulmaspri})  and for the
parametrization of radiative force we used generic parameters obtained by
Pauldrach et al. (\cite{pakupul}). The value of the microturbulent velocity was
selected to obtain smooth wind profiles. Its value does not much influence the
results obtained.

Comparisons of data from hydrodynamical simulations and from inverted data are
given in
Figs.\ref{spot05}--\ref{spot5} for spots with relative amplitudes $A=0.05,\,
0.2,\, 0.5$. We will discuss the reliability of each inverted
relation separately.

\begin{table}[hbt]
\centering
\caption{Stellar and wind parameters of HD 64760}
\label{hdpar}
\begin{tabular}{cc}
\hline
Parameter & Value \\
\hline
$R_*$ & $22\,R_\odot$ \\
$V_\mathrm{rot}$ & $238\,\mathrm{km}\,\mathrm{s}^{-1}$ \\
$T_\mathrm{eff}$ &$23\,100\,\mathrm{K}$\\
$M$ & $20\,M_\odot$\\
$L$ & $1.2\,10^5\,L_\odot$\\
$\alpha$ & $0.650$ \\
$k$ & $0.050$ \\
$\delta$ & $0.28$\\
$v_\mathrm{turb}$ & $50\,\mathrm{km}\,\mathrm{s}^{-1}$\\
\hline
\end{tabular}
\end{table}

Apparently, the inversion of the mean velocity law $v_r(r)$ is very good
(middle left panels).
Clearly, because during the inversion process we
assumed that the radial wind velocity is independent of azimuth, it cannot describe
the azimuthal variations of wind velocity caused by the outward motion of an
Abbott
kink. However, it is possible to obtain a very good inference of the 
$\phi$-averaged
velocity law, at least for low-amplitude perturbations. The main reason is that
the inversion method used is able to find the velocity law even for a
nonvariable wind via the inversion of the unperturbed absorption profile.

On the other hand the interpretation of the inverted rotational law is more
complicated. The optical
depth variations of the line profiles calculated using hydrodynamical
simulations are caused mainly by the velocity variations due to the Abbott kink. Thus,
we might expect that the inversion method would trace the movement of the
Abbott kink and
not the movement of real fluid parcels. To test this consideration, we calculated
the radial variation of the azimuth of onset for the Abbott kink relative to the stellar
spot. We define the onset of the Abbott kink as the point with the minimal
radius
where $\partial v_r/\partial r<0$.
We have compared the radial variations of the azimuth of the onset for the Abbott
kink and of the azimuth of the fluid parcel (calculated for
constant $\phi=0.4$; results for other other azimuths are very similar) with the
radial variation infered from inversion. These results are also plotted 
in the upper-right panels of Figs.~\ref{spot05}--\ref{spot5}.
Inspecting 
Figs.~\ref{spot05}--\ref{spot5} 
we
conclude that the `density' locus $\Delta\phi(r)$ found by the
inversion method in reality always traces the location of an Abbott kink
and in fact does so with a
high degree of reliability in those parts of the stellar wind 
where the Abbott kink is already formed.

We plotted also (lower panels of Figs.\ref{spot05}--\ref{spot5})
the relative azimuthal variation of the relative CIR
density profile $F(r,\phi)$ obtained using the kinematic inversion method and 
the actual density profile calculated using hydrodynamical simulations
(plotted relatively to the unperturbed wind model). Their comparison shows that
the inversion method is able to infer density profiles well only relatively near to the
stellar surface, where the optical depth variations are not yet dominated by the
variations of the velocity gradient. This is not the case in the outer parts of the
wind where both density profiles are clearly different.  This is caused by
the formation of the Abbott kink, which enhances the optical depth. Note
however that the
spot azimuth and  its width at the stellar surface can be  relatively well
inferred from the kinematic inversion method. 

The fact
that the main contribution to the DAC optical depth is given by the
moving Abbott kink also does not allow us to invert to get
correct wind densities. As 
discussed in Sect.~\ref{kininfor} the original method of Brown et al.
(\cite{brobaros}) assumes that 
the relative variations
of wind density profile in
azimuth are the same throughout the stellar wind. This also imposes limitations
on the azimuthal variations of DAC optical depths. The situation is
different with the azimuthal optical depth variations calculated with the data
from simulations, since the main contribution to the DAC optical depth in this
case is given by the moving Abbott kink. However, it is possible to slightly
modify the original Brown et al. (\cite{brobaros})  method to obtain the effective
wind density that produces the DAC from simulations. Clearly, this effective
density does not trace the real wind matter density, but rather the optical depth
variations, which are mainly given by the variation of velocity gradient due to
the 
Abbott
kink (see Eq.(\ref{sobap})). 
This situation is demonstrated in lower panels of
Figs.\ref{spot05}--\ref{spot5} where azimuthal density variations from
simulations and from inversion are compared. Apparently, relatively near the
stellar surface, the variations of velocity gradient are lower, and the optical
depth variations are influenced mostly by density variations. Thus, it is
possible to infer relatively well the correct location of a footpoint of a spot at the
stellar surface and its geometrical width. However, in the outer parts of the
stellar wind, where the Abbott kink is already formed, the contribution of
velocity gradient variations to the optical depth variations is dominant.
Thus, the inversion method in fact traces the movement of the Abbott kink. This can
be seen from upper left panels of Figs.\ref{spot05}--\ref{spot5} where density
variations obtained from inversion are compared with the location of the onset and
end of Abbott kink (defined as the radius where velocity gradient changes its sign
from positive to negative and from negative to positive respectively).


\section{Discussion}

We applied the kinematic inversion method of Brown et al. (\cite{brobaros}) to DAC
profiles calculated using hydrodynamical CIR models. We compared CIR structure
inferred by kinematic inversion with the actual simulated one, as an
important test of reliability of the  inversion method.

Although the kinematic inversion method is not able to calculate azimuthal
variation of radial velocities, 
we showed that it gives reliable results for mean wind
radial velocities even for relatively bright surface spots. However, because the
highest optical depths of DACs in the numerical simulations arise mainly from
spiralling velocity plateaux (Abbott kinks) and not from density enhancements,
the inversion method does not trace the movement of real fluid parcels, but of
an Abbott kink. Thus, it is possible from DAC data to trace the
spiral law of Abbott kinks propagating in the perturbed stellar wind. Near the
stellar surface the
spiral law of these patterns is inferred with a high degree of accuracy.
This is with quantitative accuracy and gives a general
indication of where absorbing matter lies.
Finally, it is not possible to invert to get a reliable density profile of the CIR,
due to the above mentioned dependency of optical depth variations on the
occurence of velocity plateaux, and due to the non-monotonic velocity law.
Clearly, density variations are always
overestimated. The density profile has the correct amplitude only very near to the
stellar surface, where the velocity gradient variations do not significantly
influence the optical depth. However, it is possible to obtain the correct
azimuth of the CIR footpoint  and its extent at the stellar surface.

Note that we do not require that the wind variability manifested by the presence of DACs in 
the wind profiles of hot stars be caused by hot spots at the stellar surface.
The hot spot model provides merely a convenient base perturbation for the
generation of CIRs. The DAC profile calculated using this CIR model predicts all
basic properties of observed DACs. Thus, it provides an appropriate cornerstone for
any technique aiming to infer real properties of the material causing DACs. We can
easily test which properties of CIRs can be inferred from the
observed DAC spectra and which cannot. We conclude that kinematic inversion method is
able to infer some useful information on several properties of the 
material causing the DACs
and thus can be potentially used  for observational tests of future  DAC theories.

\begin{acknowledgements}
This research has made use of NASA's Astrophysics Data System.
This work was supported by a UK PPARC Rolling Grant,
by Czech grants GA \v{C}R 205/01/0656 and
205/02/0445.
SPO acknowledges support from a PPARC fellowship for sabbatical research 
in the UK, and from NSF grant AST00-97983 to the University of Delaware. 
\end{acknowledgements}

\end{document}